\documentclass[graybox]{svmult}

\usepackage{mathptmx}       
\usepackage{helvet}         
\usepackage{courier}        
\usepackage{type1cm}        
\usepackage{makeidx}         
\usepackage{graphicx}        
\usepackage{multicol}        
\usepackage[bottom]{footmisc}

\usepackage{paralist}

\begin{document}

\title*{A survey of of blockchain-based solutions for Energy Industry}

\author{
Swati Megha
\and Joseph Lamptey
\and Hamza Salem
\and Manuel Mazzara
}

\authorrunning{Swati Megha et al.} 

\institute{Swati Megha, Joseph Lamptey, Hamza Salem, Manuel Mazzara \at Innopolis University, Universitetskaya str. 1, 420500 Innopolis, Russia} 
\maketitle

\thispagestyle{empty}
\pagestyle{empty}

\vspace{-20mm}

\abstract{The energy industry needs to shift to a new paradigm from its classical model of energy generation, distribution, and management. This shift is necessary to handle digitization, increased renewable energy generation, and to achieve goals of environmental sustainability. This shift has several challenges on its way and has been seen through research and development that blockchain which is one of the budding technology in this era could be suitable for addressing those challenges. This paper is aimed at the survey of all the research and development related to blockchain in the energy industry and uses a software engineering approach to categories all the existing work in several clusters such as challenges addressed, quality attribute promoted, the maturity level of the solutions, etc. This survey provides researchers in this field a well-defined categorization and insight into the existing work in this field from 3 different perspectives (challenges, quality attributes, maturity). }


\section{Introduction}
\label{sec:intro} 
Energy has been the fundamental engine of industry and society at least since the first industrial revolution \cite{Stern2011}. It has been a determinant factor which shaped society for centuries and it promises to be even more important in the 21st century at the time of the so-called Industry 4.0 \cite{Bassi2017}.

The cost and availability of energy significantly impacts the quality of 
life and state economy, it can lead to tensions between countries and compromise relationships between nations, therefore the stability of the geopolitical system \cite{geo-energy}. In order to foresee a future of political stability and wealth, it is necessary to rely on stable and clean forms of energies that could further foster economic development and peace between nations \cite{energy-peace}.

The 21st century is marked as the era of digital revolution where we are transforming every possible information in bits and bytes. A natural question to ask is how modern technology could foster a process of transformation toward a clean and stable energy supply and facilitate the transition to global wealth and stability. Furthermore, the energy industry showed to slowly adapt to changes in technology, so whatever can be proposed now will still take time to be deployed and utilized.

In this paper, we want to propose ideas on how the energy industry can adapt to the digital world and smart technologies for everyone's benefit, and we summarize the state-of-the-art after surveying the current literature. We will first describe in Section \ref{sec:challenges} what we consider the major challenges that the modern energy industry has to face today; then, in Section \ref{sec:requirements}, we will use our experience in Software Engineering and take a requirements engineering approach to determine what requirements for the current energy sector are emerging from the described challenges and which quality attributes are being promoted, according to the literature survey on energy and modern technologies that we have conducted. Section \ref{sec:taxonomy} categorizes the various  solutions according to requirements, that are then classified according to maturity in Section \ref{sec:maturity}. Section \ref{sec:conclusions} finally draws conclusions of the investigation.


\section{Challenges}
\label{sec:challenges} 
In this section, we will discuss what we consider the major challenges
that the modern energy industry has to face as we have entered and are in the middle of the world of technological advancements. The following five aspects can synthesize our current understanding:

\begin{enumerate}

\item{\textbf{\textit{Digitalization in energy sector: }}}
Like any other sector, the energy domain is also digitalizing its information and making it available over the internet for faster information exchange and user convenience.  Digitalization
is not only needed in today’s era but also it is a must to operate day to days business. In the 21st century, when almost everything is digitalized, several issues of the classic energy industry have to be addressed: low transparency, no real-time access to demand and supply, etc. At the same time, new challenges appear in the panorama, such as cyber-attacks, misuse of customer data, data manipulation threats, etc, which could adversely affect all the stakeholders involved in the energy industry therefore it is crucial to address these issues of privacy, security, transparency reliability: evolved smart grids is need of the hour. In this paper, we have discussed at least 9 different implementations and proposals published by researchers around the world to implement smarter smart grids and address the challenges of digitization in the energy industry.

\item{\textbf{\textit{Fossil Fuel Run out: }}}
There is hype around the world about shortage and unavailability of fossil fuel in near future, with the current rate of consumption around the world it is only a matter of time that our fossil fuel runs out. Researchers, scientists, big oil and gas companies, as well as national and international governments and organizations are trying to find alternatives to this energy crisis. The most obvious solution to this is a shift toward Renewable Energy (RE) i.e. from the sun, wind, water, etc. One of the notable features of such renewable sources of energy is that it can be produced in a distributed manner contrary to conventional energy. Due to the distributed nature of RE, it cannot utilize the year-old centralized infrastructure and demands for a new system for generation, management, and distribution. Also, RE is volatile and requires efficient demand-supply management to reduce loses in production and storage. A decentralized system for RE could be the possible solution and could be realized using blockchain technology. This paper discusses 6 different implementation and proposals facilitating a decentralized system for the Renewable energy market.

\item{\textbf{\textit{Centralized system not suited for RE generation: }}}
As mentioned in the above section that RE  is distributed in nature, it is not very convenient to produce them in a centralized manner like any other fossil fuel alternative.  Hence it demands a distributed and decentralized system of generation and distribution and management, here blockchain could play a major role to facilitate trading platforms that are suitable for RE, in this paper, we have included 6 proposals of energy trading with RE as its focus.

\item{\textbf{\textit{Sustainable Environment:}}}
After industrialization there was a huge increase in carbon emission and today, we have reached the stage where we need to think about reducing carbon footprints because it has started showing a negative effect on our climate, human health and seen as a threat to the existence of other living organisms. Hence climate change is real and has been addressed by international organizations like the United Nations \cite{UNsummit}, and targets of clean energy generation are being set. National and international organizations around the world are imposing strict rules and regulations of energy producers as well as on individual consumers to reduce their carbon footprints and take a huge leap towards decarbonization and environmental sustainability. As the need of the hour, we should focus to increase the share of RE generation and to bring advancement in technology to reduce energy wastage through efficient management and consumption. This paper discusses 11 such proposals and implementation with the focuses on the decarbonization requirement of our planet.
\end{enumerate}


\section{Categorization of Requirements and Quality Attributes}
\label{sec:requirements} 
The previous section identified several challenges that are necessary to address in energy industry. In this section we are taking a software engineering approach to surface the requirements of current energy sector from the challenges. We identified three broader requirements emerging from the previous analysis:
\begin{itemize}
\item \textbf{\textit{Digitalization: }} For communication, social and financial inclusion.
\item \textbf{\textit{Decarbonization: }} To address the issue of fossil fuel run out and environment sustainability and finally
\item \textbf{\textit{Decentralization: }} Which is a key for the generation and distribution of renewable energy
\end{itemize}
In requirements engineering, non-functional requirements (or Quality Attributes) are requirements that do not define specific behaviors, but are used to evaluate the performance of a system. Any attempt to address the challenges presented in this paper would then lead to promoting specific Quality Attributes. In the studied literature we identified 9 quality attributes as follows:
\begin{inparaenum}
\item Security
\item Transparency
\item Privacy
\item Reliability
\item Scalability
\item Efficiency
\item Ease of trade
\item Demand Supply balance
\item Promote green energy
\end{inparaenum}\\
\begin{figure}[htp!]
  \centering
    \includegraphics[width=10cm,height=7cm]{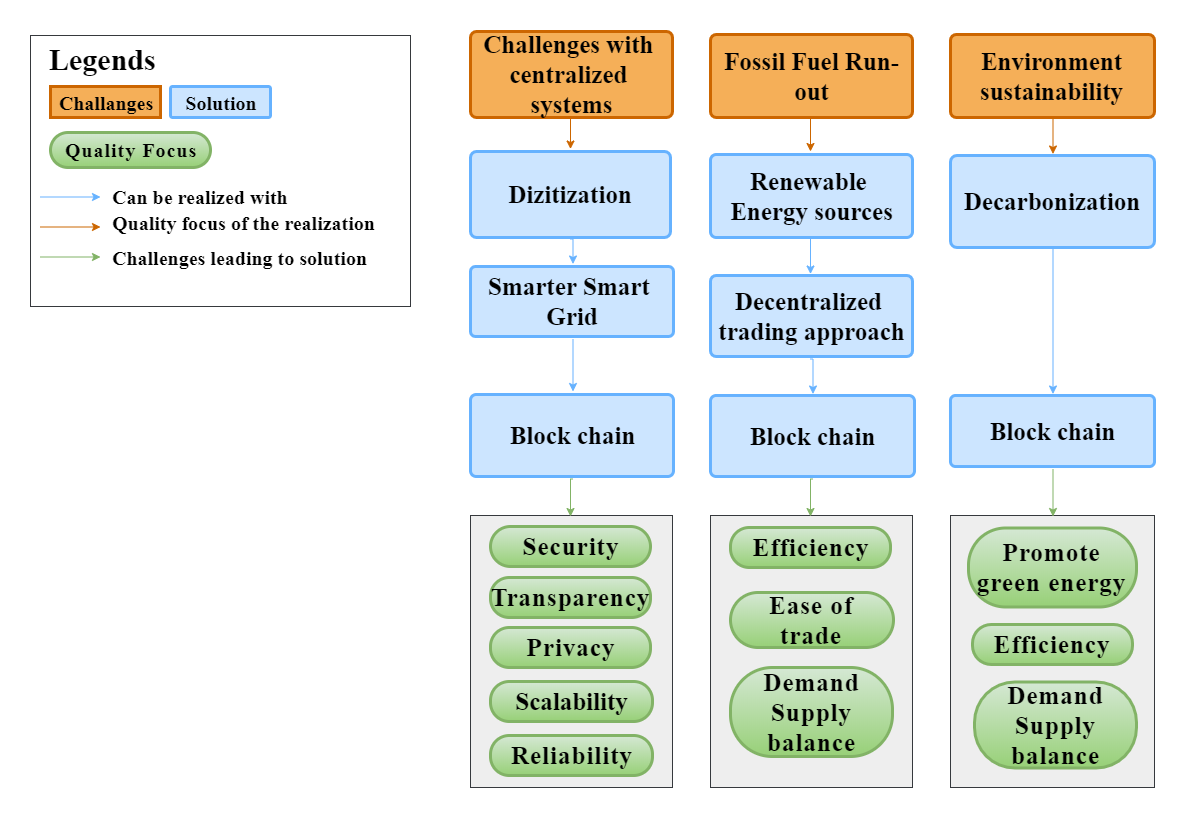}
    \caption{RequirementMapping}
\end{figure}

Digitization, decentralization and decarbonization are the major requirement of energy industry in 21st century. Blockchain which is the budding technology of this era is powerful and has the capability of addressing the requirements of energy industry as well. This paper summarises and categories the solution for energy industry implemented using block chain. The paper includes 24 different proposal and implementation of blockchain in energy industry. These papers have been categories according to the problem it addresses, the proposed solution, implementation, conclusion and the quality attributes it promotes. The paper includes 2 diagrams namely fig1 requirement mapping which shows the challenges of energy industry mapped to possible solution that could be implemented using block chain and also lists the quality attribute that could be promoted by implementing these solutions. Another important diagram is fig2 which shows well organized mapping between possible solution and implementation or proposal in different research papers and the quality focus of each research paper.\\
This paper is a well-organized survey of block chain implementation in energy industry and is helpful for researcher or interested readers in this field to get an overview, take reference from the categorization of proposal between solution category and quality attribute and for pioneer researcher to have a quick reference to different available research paper.


\section{Various solutions categorized according to requirements}
\label{sec:taxonomy} 

In this section we summarize the findings of our survey that were shortly addressed in the previous section in terms of requirements. 

\begin{enumerate}
    \item{\textbf{\textit{Digitalization in energy sector}}}\\
\textbf{Paper Title:} RP-2-Privacy-Preserving and Efficient Aggregation Based on Blockchain for Power Grid Communications in Smart Communities\cite{RP-2}.\\
\textbf{Problem Addressed:} Real-time data from user appliance is required for optimal scheduling of Smart grids but it also means disclosing user’s private information and raise the issue of privacy.\\
\textbf{Proposed Solution:} The paper proposes a privacy-preserving and efficient data aggregation scheme. Where users are divided into groups and attached with a blockchain to record data for the group members. Pseudonyms are being used to hide user identity within the group. Users are free to create as many pseudonyms and can associate their data with it. To ensure a fast authentication boom filter has been used.\\
\textbf{Conclusion:} The proposal focuses on security requirements and ensures better performance than other popular methods. It is a proposal with a simulation.\\
\textbf{Quality Focus:} Security\\\\
\textbf{Paper Title:} RP-3-GridMonitoring: Secured Sovereign Blockchain Based Monitoring on Smart Grid\cite{RP-3}.\\
\textbf{Problem Addressed:} SG readings are transmitted via the internet which generates the risk of data manipulation also with one way communication customer is unaware of consumption of appliances at their homes.\\
\textbf{Proposed Solution:}The paper proposes a sovereign blockchain technology, which provides transparency and provenance(authenticity). Realized by installing smart meters and smart contracts. The smart meter enables the user to monitor consumption and smart contract ensures no data manipulation.\\
\textbf{Quality Focus:} Transparency, Reliability\\\\
\textbf{Paper Title:} RP-4- Privacy-preserving Energy Trading Using Consortium Blockchain in Smart Grid\cite{RP-4}.\\
\textbf{Problem Addressed:} Blockchain records trading info and has the risk of attacks via data mining algorithm and misuse user’s data more threat is when the geographical location is nearby.\\
\textbf{Proposed Solution:}Proposes consortium blockchain-oriented to solve privacy leakage without compromising trading features. Implemented using the account mapping technique which hides user data hence avoids direct data access. The account is created based on the user’s usage.\\
\textbf{Conclusion:} Investigated privacy issues in blockchain-based neighboring energy trading system on the locality of 20 neighbors.\\
\textbf{Quality Focus:} Privacy\\\\\
\textbf{Paper Title:} RP-5- Blockchain: A Path to Grid Modernization and Cyber Resiliency\cite{RP-5}.\\
\textbf{Problem Addressed:} Current power grid infrastructure cannot prevent cyber-attacks on DERs. Cyber vulnerabilities and interoperability is a major challenge in the energy sector.\\
\textbf{Proposed Solution:}Paper explores applications of blockchain and smart contract to enhance smart grid cyber resiliency and secure transactive energy applications.\\
\textbf{Conclusion:} Generally, testbeds used to test such systems are not as complex as power grid’s PNNL’s B2G testbed and integrated Transactive Campus provide a unique combination of live telemetry and real-time data to simulate the power grid and improve the state of the art of blockchain security technology to create a more resilient grid.\\
\textbf{Quality Focus:} Security\\\\
\textbf{Paper Title:} RP-9- A Blockchain Model for Fair Data Sharing in Deregulated Smart Grids
\textbf{Problem Addressed:} User data important for service provider for efficient service, however, it raises privacy concerns\cite{RP-9}.\\
\textbf{Proposed Solution:}Solution proposed is to fairly compensate sues for sharing their data. Implemented via blockchain and the concept of differential privacy. Derives reputation scores through the Pagerank mechanism.\\
\textbf{Conclusion:} Shows how anonymity as a measure to conceal information to achieve a minimum leakage. Provides simulation to shows change customers behavior towards privacy with the proposal.\\
\textbf{Quality Focus:} Privacy\\\\
\textbf{Paper Title:} RP-10- Distributed Blockchain-Based Data Protection Framework for Modern Power Systems against Cyber Attacks\cite{RP-10}.\\
\textbf{Problem Addressed:} Cybersecurity for the robustness of modern power systems.\\
\textbf{Proposed Solution:}Proposes a distributed blockchain-based protection framework that focuses on enhancing the self-defensive capability of modern power systems against cyber-attacks. Meters are used as a node in distributed network and meter measurements are blocks.\\
\textbf{Conclusion:} Effectiveness of proposal demonstrated via simulation.\\
\textbf{Quality Focus:} Security\\\\
\textbf{Paper Title:} RP-19-Blockchain for Power Grids\cite{RP-19}.\\
\textbf{Problem Addressed:} Information exchange in power grids in a secure, controlled, monitored and efficient manner.\\
\textbf{Proposed Solution:}Proposes a Hyperledger Fabric implementation with permission network, power grids act as nodes that stores ledger information. Transaction validation is done via the process of consensus.\\
\textbf{Quality Focus:} Security, efficiency\\\\
\textbf{Paper Title:} RP-21-Blockchain for Smart Grid Resilience Exchanging Distributed Energy at Speed, Scale, and Security\cite{RP-21}.\\
\textbf{Problem Addressed:} Lack of security and resilience to prevent cyber-attacks on DERs, grid edge devices, and associated electricity infrastructure.\\
\textbf{Proposed Solution:}This paper explores the application of blockchain and smart contracts to improve smart grid cyber resiliency and secure transactive energy applications.\\
\textbf{Quality Focus:} Cyber security\\\\
\textbf{Paper Title:} RP-1-A Blockchain-Based Access Control Scheme for Smart Grids.\cite{RP-1}\\
\textbf{Problem Addressed:} At present power grids have Centralized Access control schemes (ACS) which have centralized or third-party authentication generating the potential risk of data loss and manipulation as well as key escrow problem of the untrusted third parties.\\
\textbf{Proposed Solution:}The paper proposes blockchain-based decentralized access control system, the system is based on consortium blockchain, proposes consensus algorithm for the selection of private key generator (PKG) in smart power grids with incentive and penalty mechanism, the PKG here is reusable and could be used for many scenarios.\\
\textbf{Conclusion:} The proposed scheme focuses on satisfying security requirements of the power grid, on solving the key escrow problem, making the user more involved in the daily management of smart grids and reusability of PKG reduces communication costs and suits multi-environments in smart grids.\\
\textbf{Quality Focus:} Reliability\\\\
\textbf{Paper Title:} RP-16-Security and Privacy in Decentralized Energy Trading through Multi-Signatures, Blockchain, and Anonymous Messaging Streams\cite{RP-16}.\\
\textbf{Problem Addressed:} Security and scalability concerns of centralized trading that involves the third party.\\
\textbf{Proposed Solution:} Implemented PriWatt a trustless decentralized token-based energy trading system with anonymous user communication and energy ownership in SG distributed smart contracts. Implemented using blockchain technology, multi signatures, and anonymous encrypted message propagation streams. Uses future energy demand prediction to balance demand and supply.\\
\textbf{Conclusion:} Implemented two algorithms namely Energy Trading Algorithm and Micropayment.\\
\textbf{Quality Focus:} Privacy and Security\\\\
\item{\textbf{\textit{Fossil Fuel Run out and Centralized system not suited for RE generation }}}\\
\textbf{Paper Title:} RP-12-How Blockchain Could Improve Fraud Detection in Power Distribution Grid\cite{RP-12}.\\
\textbf{Problem Addressed:} Addressing electrical fraud that results in large losses while Power distribution.\\
\textbf{Proposed Solution:}Proposes a system to detect fraud implemented using blockchain which stores data collected by the WSN that monitors the power distribution grid then construct directed acyclic graph (DAG) from this data and then apply a new clustering algorithm to detect fraud.\\
\textbf{Quality Focus:} Efficiency\\\\
\textbf{Paper Title:} RP-11- Blockchain-based electricity trading with Digital grid router.\\
\textbf{Problem Addressed:} Challenge of providing Secured and decentralized control to peer-to-peer exchanges\cite{RP-11}.\\
\textbf{Proposed Solution:}Proposes an Ethereum Blockchain-based electricity trading system incorporates a digital router as a platform which consists of back-to-back bi-directional digital inverters with software-based control enables/facilitate power exchange.\\
\textbf{Conclusion:} Shows simulation that individual exchange can improve the efficiency of energy compared with the no-exchange case in microgrid case.\\
\textbf{Quality Focus:} Efficiency\\\\
\textbf{Paper Title:} RP-13- A Blockchain-Based Energy Trading Platform for Smart Homes in a Microgrid\cite{RP-13}.\\
\textbf{Problem Addressed:} Security in renewable energy trading.\\
\textbf{Proposed Solution:}Proposes trading platform based on Ethereum smart contract focuses on secured trading without the third part in microgrids. Trading transactions are automatic.\\
\textbf{Conclusion:} Implemented simple scenario for two nodes that shows transactions are automatic and no possibility of forge because of blockchain and smart contracts.\\
\textbf{Quality Focus:} Security\\\\
\textbf{Paper Title:} RP-17-Smarter City Smart Energy Grid based on Blockchain Technology\cite{RP-17}.\\
\textbf{Problem Addressed:} Investigate SEGs and blockchain in urban smart cities.\\
\textbf{Proposed Solution:} Implemented a Smart Energy Grid based on Blockchain Technology \& Blockchain\_SEG Application a mobile application for grid information exchange and buy and sell energy digitally at fingertips. Implemented using blockchain granting ledger.\\
\textbf{Conclusion:} Enhances the quality of life at smart cities at the same time provides efficient and flexible energy trading.\\
\textbf{Quality Focus:} Efficiency, ease of trade, Eco-Sustainable vision\\\\
\textbf{Paper Title:} RP-23-Crypto-Trading Blockchain-oriented energy market\cite{RP-23}.\\
\textbf{Problem Addressed:} Electricity management in smart grids.\\
\textbf{Proposed Solution:}Modular blockchain-based software called crypto trading project launched with a Fintech company. It extends feature of a cryptocurrency exchange to RE energy market including garobo-advisor suggesting prosumers the bestselling strategy.\\
\textbf{Quality Focus:} Ease of trade\\\\
\textbf{Paper Title:} RP-24-Enabling peer-to-peer electricity trading\cite{RP-24}.\\
\textbf{Problem Addressed:} Platform required for peer-to-peer (P2P) electricity trading between households what has emerged with distributed energy sources.\\
\textbf{Proposed Solution:}Proposes a peer-to-peer (P2P) electricity trading platform. Implemented using blockchain and hence provides decentralized, distributed ledger in which all transactions are immutably recorded relies on cryptography to ensure transactions are secure, authenticated and accurate. Trading in such system is traceable and auditable hence produces accurate bills.\\
\textbf{Quality Focus:} Ease of trade\\\\
\item{\textbf{\textit{Sustainable Environment}}}\\
\textbf{Paper Title:} RP-6-M2M Blockchain: The Case of Demand Side Management of Smart Grid\cite{RP-6}.\\
\textbf{Problem Addressed:} Addresses techniques for demand-side management of the smart grid.\\
\textbf{Proposed Solution:}Present a method to manage the demand and transaction of the grid using blockchain and smart contract for independent maintenance and management of transaction information as well as the automatic transfer of funds in the whole network node.\\
\textbf{Conclusion:} Shows that we can actively adjust the power generation trading with each other over a blockchain. and contributes to proof-of-concept implementation provides scenarios verifies the feasibility of the method.\\
\textbf{Quality Focus:} Demand Side Management\\\\
\textbf{Paper Title:} RP-8- Energy Demand Side Management within micro-grid networks enhanced by blockchain\cite{RP-8}.\\
\textbf{Problem Addressed:} Challenges with managing supply and demand of energy.\\
\textbf{Proposed Solution:}Proposes game-theoretic model for Demand Side Management (DSM) model with storage component.\\
\textbf{Conclusion:} Model shows a reduction in Peak-to-Average ratio and smoothens the dips in the load profile caused by supply constraints.\\
\textbf{Quality Focus:} Demand Side Management\\\\
\textbf{Paper Title:} RP-14-A blockchain-based smart grid towards sustainable local energy\cite{RP-14}.\\
\textbf{Problem Addressed:} Centralized energy market used for NRE is not suitable for RES which is volatile, requires distributed generation and real-time mechanism to react in the wholesale market.\\
\textbf{Proposed Solution:} Managing decentralized generation through digitalization. The proposed new market mechanism called Local Energy Market (LEM). It was proposed by the European commission in 2016. Implemented using private and permission blockchain which doesn’t require any central authority. LEM is limited to the geographical or social community and balance demand and supply locally. The mathematical formula is developed for accurate calculation of demand and supply and is being forecasted.\\
\textbf{Conclusion:} Implemented online mechanism matching supply and demand, reduce trading complexity by forecasting consumption and generation, the system does not involve any intermediaries. The system is built for small geographical communities using private and permission blockchain which ensure greater security.\\
\textbf{Quality Focus:} Demand Supply balance, Security, Efficiency, Ease of trade\\\\
\textbf{Paper Title:} RP-18-Blockchain Based Decentralized Management of Demand Response Programs in Smart Energy Grids.\\
\textbf{Problem Addressed:} Issues of demand-supply management\cite{RP-18}.\\
\textbf{Proposed Solution:}Proposes a blockchain-based distributed ledger the ledger stores the information of prosumers collected from the internet of things in a tamperproof manner and self-enforce smart contracts. prosumers get incentives or penalties for balancing demand-supply. Consensus-based validation is used for financial settlements.\\
\textbf{Conclusion:} The prototype was implemented in Ethereum to validate and test the blockchain-based decentralized management using energy traces of UK building datasets. Results show that the grid is capable of balancing energy demand in near real-time.\\
\textbf{Quality Focus:} Demand Supply balance, Ease of trade\\\\
\textbf{Paper Title:} RP-22-Designing micro-grid energy markets A case study The Brooklyn Microgrid-new\cite{RP-22}.\\
\textbf{Problem Addressed:} Demand supply balance in distributed RES generation which by nature volatile.\\
\textbf{Proposed Solution:}Proposes self-trading within the community and also keeping the profit within the community, blockchain-based decentralized energy market without central intermediaries.\\
\textbf{Quality Focus:} Demand Supply balance, Ease of trade\\\\
\textbf{Paper Title:} RP-7- Feather forking as a positive force: incentivizing green energy production in a blockchain-based smart grid.\\
\textbf{Problem Addressed:} Increased use of non-renewable energy\cite{RP-7}.\\
\textbf{Proposed Solution:}Proposes an SG architecture(distributed) incorporating blockchain which focuses on discouraging production and distribution of non-renewable energy. An ethical mechanism is being introduced inside blockchain to incentivize the green energy production the paper discusses the feather forking attack and formalize it for green energy.\\
\textbf{Conclusion:} The proposed techniques seems promising for ethical smart grid system where all stakeholders collaborate for a greener environment.\\
\textbf{Quality Focus:} Green Energy \\\\
\textbf{Paper Title:} RP-15-NRGcoin\_virtual currency for trading of renewable energy in smartgrids\cite{RP-15}.\\
\textbf{Problem Addressed:} Implement Green energy. Distributes energy market for RES generally rely on the day-ahead prediction of demand and supply for trading. This forecast needs to be near to accurate and also prosumer have to develop a trading strategy to maximize profit.\\
\textbf{Proposed Solution:} Proposes new digital currency for energy trading called NRGcoin. Along with a novel trading paradigm. Balances demand-supply by providing incentives to prosumer to motivate them to regulate production and consumption out of self-interest. NRG coin is different from Bitcoin as it is generated by injecting energy into the grid whereas bitcoin is generated by spending energy on computational power. Value of coin determined via open currency exchange market and can be exchanged at any time by prosumers.\\
\textbf{Quality Focus:} Demand Supply balance, Ease of trade, Scalability, flexibility\\\\\
\textbf{Paper Title:} RP-20-Bitcoin‐Based Decentralized Carbon Emissions Trading Infrastructure Model\cite{RP-20}.\\
\textbf{Problem Addressed:} Reduce carbon emission and maintain privacy while energy trading.\\
\textbf{Proposed Solution:}Proposes a system-of-systems architecture model for a Decentralized Carbon Emissions Trading Infrastructure (D-CETI). stakeholders do not have to reveal their identities to be able to participate in the marketplace.\\
\textbf{Proposed Solution:}
D-CETI is designed generate carbon credits, trade and manage them. It has inherited the bitcoin mining process based on cryptographic proof of work, dataset is distributed and public and provides flexible way of participating in distributed network there is no central bank or authority to manage carbon emission credits. It is a high privacy system focused on reducing carbon emission.\\
\textbf{Quality Focus:} Privacy, Carbon emission reduction, green energy.\\\\\\\\

\end{enumerate}
\begin{figure}[htp!]
\includegraphics[width=15cm,height=11cm]{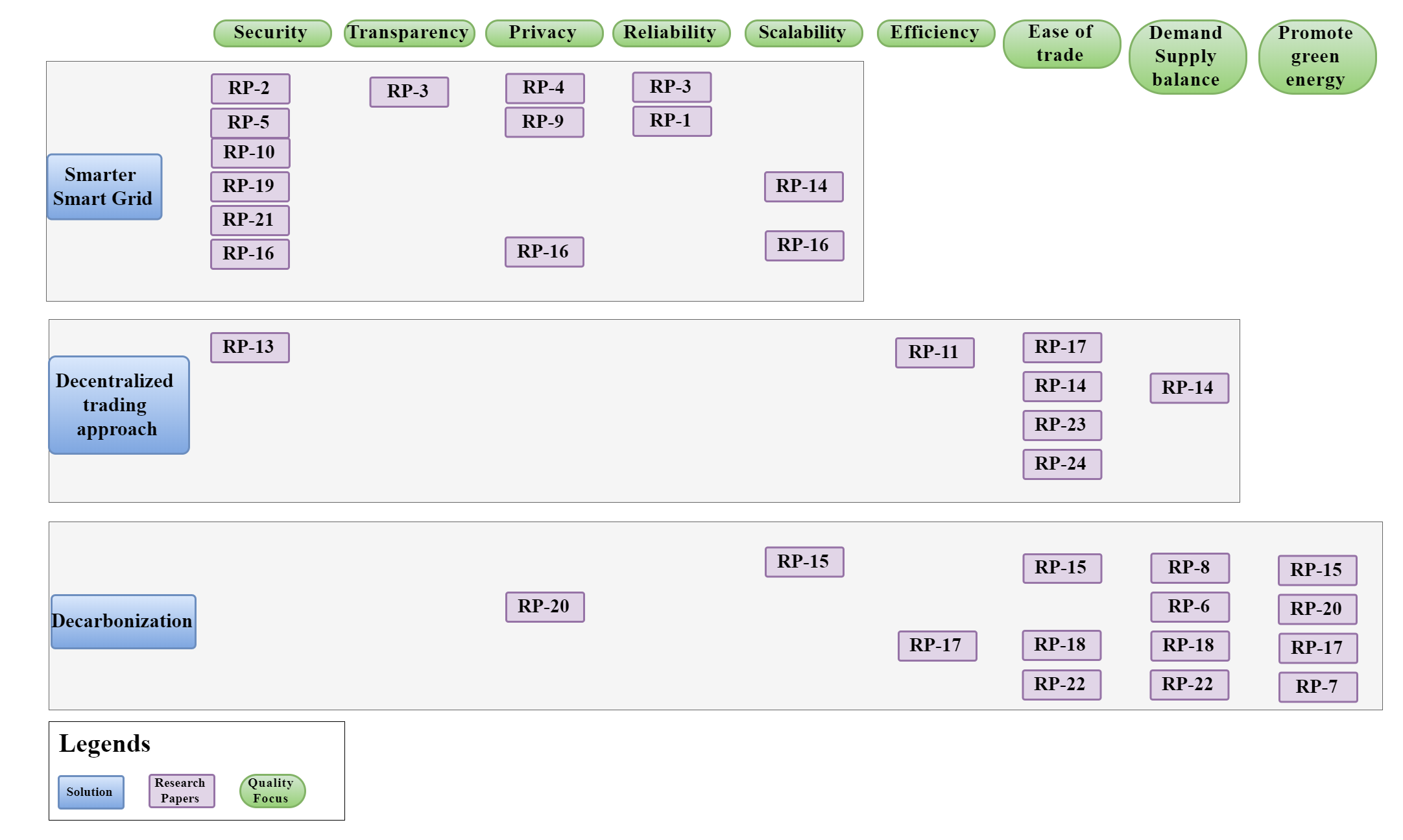}
    \caption{PaperMapping}
    \label{fig:figure}
\end{figure}


\section{Maturity}
\label{sec:maturity} 

The different solutions we have surveyed can also be categorized according to an orthogonal dimension, meaning how solid they are considering a range from idea to profitable product. We will use here the following scale:

\begin{itemize}
\item Idea: nothing has been done apart from a proposal 
\item Concept: the idea has been structured and conceptualized
\item Prototype: there is a minimal working system 
\item Product: there is a fully functional system that could be commercialized
\item Profit: the system is able to generate profit
\end{itemize}
Most of the survey paper included here could be categorized into the top 4 categories namely idea, concept, prototype and product. RP-11\cite{RP-11}, RP-23\cite{RP-23}, RP-24\cite{RP-24} is just an idea of blockchain-based electricity trading system. RP-15\cite{RP-15} is a just idea/concept of new currency to trade renewable energy. RP-5 \cite{RP-5} and Rp-6 \cite{RP-6}, RP-21\cite{RP-21} are just exploring the use of blockchain to improve cybersecurity and demand-side management respectively in smart grids and do not propose any solution (Exploring several ideas). Whereas RP-22\cite{RP-22} presents a state-of-art literature review of the combination of microgrid and blockchain and presents the key findings of the review. Altogether we categorize RP-11, RP-23, RP-24, RP-5, RP-15, RP-21, RP-6, RP-22 as an idea.\\
Rp-4\cite{RP-4} proposes a Privacy-preserving Energy Trading system and compare and investigate it with other existing models to conclude that their model is better. RP-7 \cite{RP-7}is all about proposing a blockchain-based architecture aimed at regulation of energy production and distribution. RP-8\cite{RP-8} proposes a demand-side management model and performed a case study to demonstrate the value of their model. RP-12\cite{RP-12} proposes a model for fraud detection in power grids and suggest some clustering algorithm without any simulation data or results. RP-13\cite{RP-13} proposes an energy trading system for renewable energy and present software code of trading process using smart contracts. RP-18\cite{RP-18} proposes a solution for demand-side management and developed a prototype and showed results of implementation. Altogether RP-4, RP-7, RP-8, RP-12, RP-13, and RP-18 could be categorized as a proposal.\\
Paper RP-1\cite{RP-1}, RP-2\cite{RP-2}, RP-9\cite{RP-9}, RP-10\cite{RP-10}, RP-16\cite{RP-15}, RP-19\cite{RP-19}, RP-20\cite{RP-20} have a concrete design for their system have already implemented algorithms or software code and have done quality tests such as scalability, transparency, security, performance etc and presented analysis and simulation results. So all together we can categorize these papers into an existing prototype.\\
Whereas as RP-3\cite{RP-3} has proposed design algorithms and has also implemented a sovereign blockchain-based system on the smart grid and have shown real results for the improved quality attributes such as transparency. RP-14\cite{RP-14} presents a comprehensive concept, market design, and simulation of a local energy market between 100 residential households. RP-17\cite{RP-17} proposes a complex architecture for the implementation of SGs using blockchain and implemented a mobile application for information exchange among prosumers hence fall in the category of product. 

\vspace{-5mm}

\section{Conclusion}
\label{sec:conclusions} 

This paper has shown how Blockchain technologies can address major challenges of emerge industry with the help of blockchain technology:
\begin{inparaenum}
\item Digitalization
\item Decarbonization
\item Decentralization
\end{inparaenum}

The paper survey and categorize challenges of the energy industry in light of the drift towards industry 4.0, shows a general readiness and willingness to embrace solutions from blockchain technology. The research papers included in this survey embodies the perspectives of researchers, experts, analyst, businesses and environmentalists. It attempts to highlights and emphasizes the salient works that have been simulated, published or documented. It provides illumination for the next phase of work and contemporary research ongoing as well as challenges for future work. To arrive at these, a concise and analytical approach was adopted from the software engineering fraternity to not only craftily simplify the volume of work in this field but also offer a template suitable in documenting and discussing the trends, challenges, and solution in the energy sector as a strategic component of the industrial revolution. The paper categorizes the research work according to challenge it is addressing quality attributes it promotes and at last also specify the maturity of each paper. Classification of maturity shows that there is immense work to be done in order to commercialize the existing proposals, as most of them are either idea, concept or prototype. All this categorization together could be helpful to beginner researcher or expert into this field to look at the existing work in a structured manner and can give an insight into the paper form different categorized prospective.\\
Without a doubt, the research works surveyed, depict that digitalization is very instrumental in the modern energy sector just as decentralization and decarbonization. It is worth noting that unlike traditional energy sources, industry 4.0 relies on renewable sources of energy that are deemed distributed or decentralized by nature. This further accentuates the notion to exploit state of the art technology such as a Blockchain network in distributing it after harnessing it within the localities. It also opens the opportunity for individuals to participate in the production and sale of energy. More importantly, it enables authorities to maneuver local challenges of energy supply in the face of difficulties with the national grid. This certainly encompasses the protection and preservation of the environment and addressing climate change brought about by the orthodox energy exploration. There are desirable attributes such as reliability, security, scalability, efficiency, and mobility among others that constitute some of the tenets we used to assess the renewable energy types. We find this to be very useful and strongly recommend the adoption of this approach. However, quality attribute scenarios are not easily distinguished from the functionality of the energy system. This is a caution that a designer and research should consider. The suggested taxonomy and the orthogonal criteria represent a conceptual framework that could be effectively applied to different areas. We are planning to extend the research in the field focusing on broader solutions other than the only Blockchain.


\bibliographystyle{plain}
\bibliography{references}
\end{document}